\documentclass[aps,superscriptaddress,twocolumn,preprintnumbers,floatfix,showpacs,prx]{revtex4-1}

\usepackage{amsmath}
\usepackage{physics}
\usepackage{amsfonts}
\usepackage{amssymb}
\usepackage{graphicx}
\usepackage{caption}
\usepackage{subcaption}
\usepackage[utf8x]{inputenc} 

\begin{document}

\title{Non-Hermitian topology in molecules: Prediction of fractional quantum number}
\author{Jimin Li}
\email{jl939@cam.ac.uk}
\affiliation{Department of Chemistry, University of Cambridge, Lensfield Road,
Cambridge, CB2 1EW, UK.}
\date{\today}
\begin{abstract}
    We give a simple toy model to study a famous Jahn-Teller type molecule. Finite lifetime due to non-adiabatic coupling and finite temperature effect results in the effective Hamiltonian to be non-Hermitian. This effect pulls a conical intersection into a pair of connected Weyl points, bridged by a Fermi arc. The length of the Fermi arc depends on the strength of non-hermicity. This is a unique feature of non-Hermitian topology with no Hermitian analogue. We predict the existence of Weyl points in molecules which cause anomalous Jahn-Teller effects and fractional quantum number.
\end{abstract}
\maketitle
\section{Introduction}

Over the past decades, the topology of physical systems has led to the discovery of quantum hall effect, topological insulator and many other new states of matters\cite{fu2007topological}. One attractive feature of a topological system is the robustness against perturbations. The robustness and classification of matters are understood by identifying the topological invariant within the framework of topological band theory\cite{armitage2018weyl}. Interesting phenomena in this fruitful field has its origin in the geometric phase. This was first discovered in molecular systems where the appearance of conical intersections(CI) plays an important role in Jahn-Teller effect and nuclear dynamics\cite{mead1979determination}. CI influences the interference pattern of a molecular scattering experiment as well as the vibrational spectrum\cite{yuan2018observation}\cite{von1998unambiguous}. In recent years, non-Hermitian(NH) quantum mechanics has been studied extensively, such as laser and photonics\cite{zhen2015spawning}. A unique feature of non-hermitian is the existence of exceptional points(EP) where both eigenvalues and eigenstates coalesce. There has been an intensive theoretical study on NH models in topological systems, e.g. non-Hermitian skin effect and bulk Fermi arc\cite{yoshida2019symmetry}\cite{yoshida2018non}\cite{kozii2017non}. Experimental photonics studies have realised some of the predicted phenomena such as bulk Fermi arc and exceptional ring\cite{cerjan2019experimental}\cite{zhou2018observation}. Sadly, barely any studies have done in molecules. In materials and molecules, the true many-body Hamiltonian is always hermitian. NH Hamiltonian occurs as an effective one-body quasiparticle Hamiltonian. It can emerge as a result of non-zero temperature effect, nonadiabatic coupling, strong correlations or coupling to a continuum\cite{nagai2020dmft}\cite{michishita2020equivalence}\cite{papaj2019nodal}. The use of NH is ubiquitous in self-energy corrections in strongly correlated electrons where the imaginary part represents the lifetime of quasiparticle. However, the NH topology has not been considered until a very recent study\cite{shen2018topological}\cite{bergholtz2019exceptional}. In this work, we introduce a simple toy model to demonstrate a striking consequence of exceptional points in a well-studied Jahn-Teller molecule.

\section{Conventional Geometric Phase}
Under the Born-Oppenheimer(BO) approximation, the molecular Hamiltonian parametrically depends on the nuclei coordinates $\mathbf{R}$. Geometric phase can be defined by 
\begin{equation}
\gamma_{n} = i\oint_{C} \bra{\psi_{n}}\ket{\nabla_{\mathbf{R}} \psi_{n}}d\mathbf{R}
\end{equation}
Where $\psi$ is an electronic BO wave function. It can be shown that for a hermitian Hamiltonian\cite{mead1979determination}, $e^{i\gamma_{n}}=-1$ if $C$ encircles a CI and $0$ otherwise. Namely, using a real BO electronic wavefunctions, the nuclear wavefunction must changes sign for any path that encloses a CI once. 

Consider an $X_{3}$ molecule with $D_{3h}$ symmetry. Using Herzberg's notation\cite{herzberg1945molecular}, let $Q_{2a}$ and $Q_{2b}$ be the two degenerate normal coordinates, $\psi_{ea}$ and $\psi_{eb}$ be the degenerate electronic states. Then, define the complex normal coordinates and electrons states $Q_{2\pm} = Q_{2a} \pm iQ_{2b}$ and $\psi_{e\pm} = \psi_{ea} \pm i\psi_{eb}$ simplify the calculation. Neglecting all the quadratic terms and by a simple symmetry argument, we obtain
\begin{equation}
H_{0} = \begin{pmatrix}
E_{0}  & CQ_{2-} \\
CQ_{2+} & E_{0} 
\end{pmatrix}
\end{equation}
where $E_{0} = \bra{\psi_{e\pm}}\hat{H}_{ele}\ket{\psi_{e\pm}}$ and 
\begin{equation}
C = \frac{\partial \bra{\psi_{e\pm}}\hat{H}_{ele}\ket{\psi_{e\mp}}}{\partial Q_{2\pm}}
\end{equation}
are assumed to be non-zero constant for small vibrations. The CI provides a $\pi$-magnetic flux and the sign of electron state changes upon $2\pi$ rotation in the $Q_{2}$ parameter space. On the other hand, the total wavefunction must be single-valued. This means the vibrational wavefunctions must change sign. For an approximate rovibrational spectrum,
\begin{equation}
G_{u,j} = (u+\frac{1}{2})\omega + Aj^{2}
\end{equation} 
where $\omega$ and $u$ are the frequency and quantum number of radial vibrations. Vibronic coupling leads the new quantum number $j$. By the above argument, $j$ must be half-integer valued, which was observed in the vibrational spectrum of $Na_{3}$ by comparing the experimental peaks with ab initio peaks\cite{von1998unambiguous}.
\section{Exceptional Points}
In the studies of quasiparticle, Green function and the total Hamilnotian are the most useful concepts. Quasiparticle excitations are associated with the poles of the Green function. The later one is given by
\begin{equation}
H = H_{0} + iH_{self}(\omega)
\end{equation}
where
\begin{equation}
H_{self}(\omega) = \begin{pmatrix}
\Gamma_{1}(\omega) & \Gamma_{2}(\omega) \\
\Gamma_{3}(\omega) & \Gamma_{4}(\omega)
\end{pmatrix}
\end{equation}
The time-dependent self energy $iH_{self}$ is non-Hermitian when the lifetime of a quasiparticle is finite. In this story, it is coming from nonadiabatic coupling and finite temperature effect. If $X$ is a heavy d/f block element, e.g. Cerium in a molecule which contains delocalised electron ligands, e.g. phosphate and carboxylic anion. Then electron-electron interaction can introduce the same effect\cite{michishita2020equivalence}. In a nonadiabatic coupling calculation, TD-HF and TD-DFT can both produce imaginary energy but often been discarded\cite{li2014first}. For simplicity of the model, we expand $\Gamma$ in a series and assume $\omega$ is small. Therefore, $\Gamma$ is a non-zero constant. Furthermore, we ignore the diagonal terms since it merely shifts the levels. 
\begin{equation}
iH_{self} = i\Gamma\hat{\sigma}_{x}
\end{equation}
Direct diagonalisation gives the energy of two levels 
\begin{equation}
E_{\pm} = E_{0} \pm \sqrt{Q_{2a}^2 + Q_{2b}^2 + 2iQ_{2a}\Gamma - \Gamma^{2}}
\end{equation}

\begin{figure}[hpt]
\begin{subfigure}{0.5\textwidth}
\includegraphics[width=1\linewidth, height=5cm]{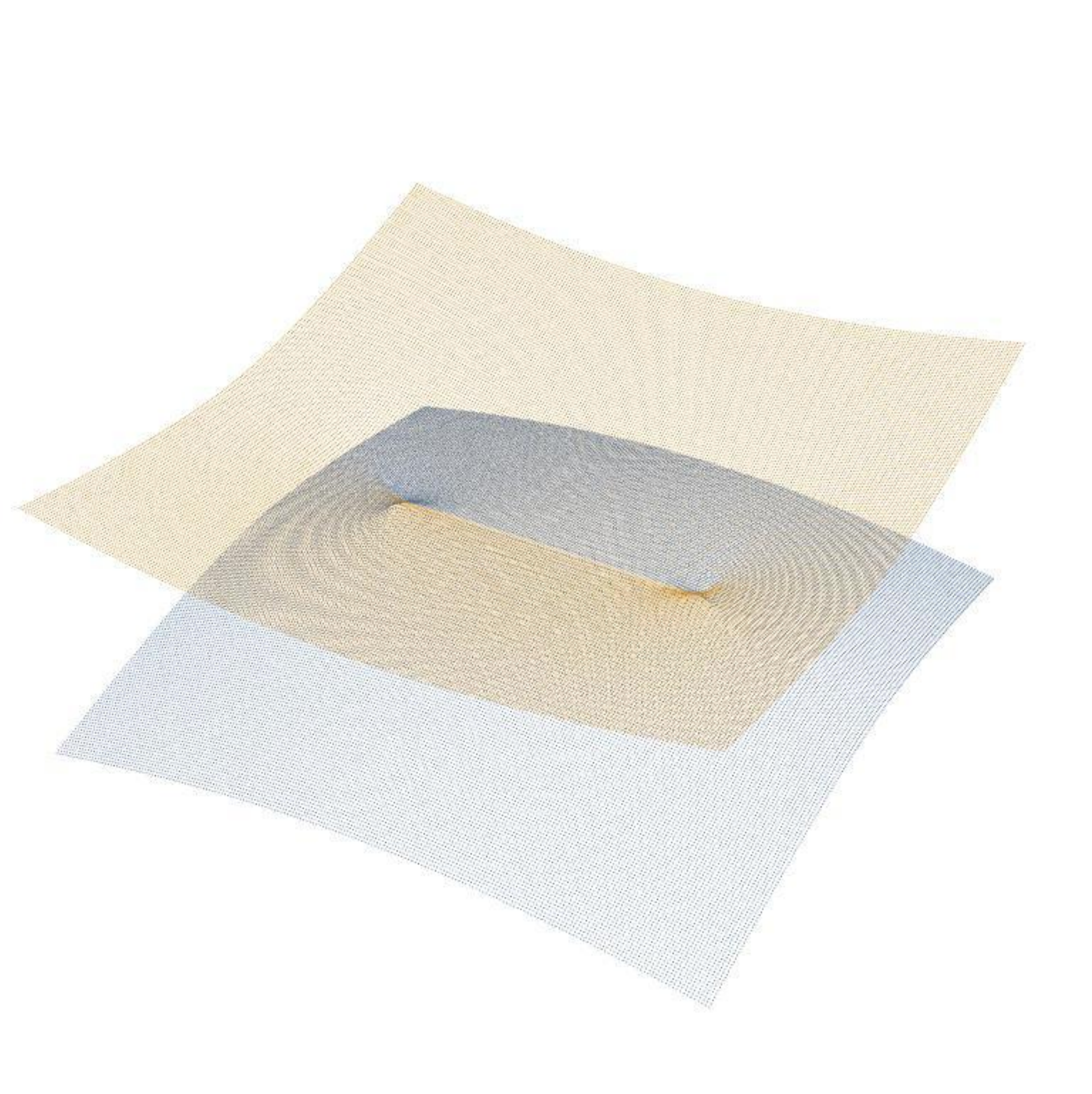} 
\caption{Real part of the eigenvalues}
\label{fig:subim1}
\end{subfigure}
\begin{subfigure}{0.5\textwidth}
\includegraphics[width=1\linewidth, height=5cm]{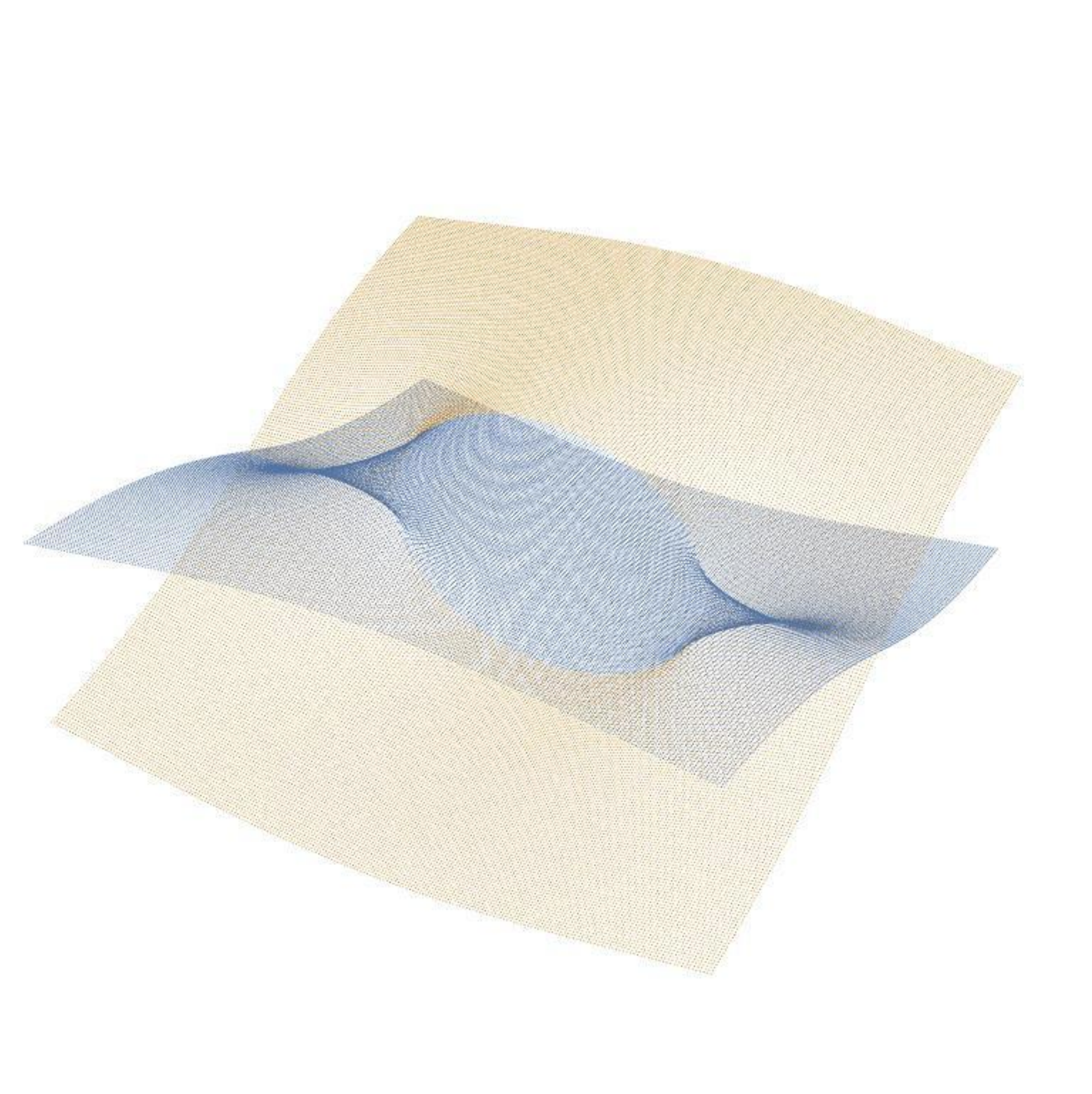}
\caption{Imaginary part of the eigenvalues}
\label{fig:subim2}
\end{subfigure}
\caption{Schematic plot of the energy spectrum using Cartesian coordinates. $x$ and $y$ directions represent the normal coordinates $Q_{2a}$ and $Q_{2b}$. $z$ direction is used for energy. }
\label{fig:image2}
\end{figure}

We get the typical square root spectrum and observe two eigenstates coalesce at the intersection. At the exceptional points, the hamiltonian is not diagonalisable, i.e. the best one can get is a Jordan block. Physically, non-hermicity split a CI into a pair of Weyl points which is connected by a Fermi arc. This Fermi arc is fundamentally different from the commonly observed surface intersection of PES in molecules. Unlike topological band theory where the parameter Bloch vector is a good quantum number. In a molecular PES, the generalised nuclei coordinates are not. CI requires a perfect match of Hamiltonian matrix elements and can be broken by arbitrary small interactions. However, the Weyl points are topologically stable under perturbation.  Each of the exceptional points is defined to be half-integer topological charge within the framework of NH topological band theory\cite{shen2018topological}. Encircling only one of them,
$\ket{E_{\pm}}$ becomes to $\ket{E_{\mp}}$ and return to $-\ket{E_{\pm}}$ after a second $2\pi$ rotation. This implies that the vibrational wavefunctions change sign after a rotation of $4\pi$. Meaning that $j$ are forbidden to be half-integer and  
quantised to be $|J| = \frac{1}{4},\frac{3}{4},\frac{5}{4}...$\cite{wuczek2002some}\cite{wilczek1984appearance}. Encircling any closed loop contains both Weyl points will change the sign of $\ket{E_{\mp}}$. When $\Gamma$ is sufficiently small, two Weyl points can be inseparable and therefore behave like a CI. The robustness of Weyl points can explain why a fragile CI exists along with a chemical reaction when higher-order perturbations and many other effects are usually not been considered. A similar scenario can occur in a CI of the ground state and excited state. Encircling one of the Weyl points in the ground state results the system in an excited state which can be achieved by a series of molecular vibrations in real space. In three and higher dimensions, NH topological band theory has predicted a number of new topological matters\cite{carlstrom2019knotted}. By the same token, similar physics can happen in molecules with higher degrees of freedom.
\section{Summary}
In this work, we studied non-hermitian $X_{3}$ systems. The non-hermiticity is a result of nonadiabatic coupling and temperature effect which split a CI into a pair of connected Weyl points. We predict that the possibility of anomalous Jahn-Teller effects and fractional quantum number $j$ in $X_{3}$. This non-hermiticity effect should be ubiquitous in molecular science since no spectrum peak is a delta function. This work is no means a complete theory but merely to show such simple model yield unexpected phenomena and many potential cases in molecules such as new new approach to solution/surface reactions and roaming mechanism.
\bibliographystyle{unsrt}
\bibliography{paper}

\end{document}